\documentclass[11pt,a4paper,usenatbib]{emulateapj}
\slugcomment{{\sc Accepted to ApJ:} 7 February, 2008. Reproduced with the permission of the AstroPhysical Journal}

\usepackage{amsmath}

\newcommand{\Msol}{\rm\,M_{\odot}}
\newcommand{\Mstellar}{\rm\,M_{*}}
\newcommand{\Mcr}{\rm\,M_{cr}}
\newcommand{\Mhalo}{\rm\,M_{halo}}
\newcommand{\lambdaobs}{\rm\,\lambda_{obs}}
\newcommand{\lambdarest}{\rm\,\lambda_{rest}}
\newcommand{\fire}{\rm\,f(IRE)}
\newcommand{\Rc}{\rm\,R_{C}}
\newcommand{\UBVRI}{\rm\,UBVR_{C}I_{C}}
\newcommand{\Gyr}{\rm\,Gyr}
\newcommand{\Msolyr}{\rm\,M_{\odot}\,yr^{-1}}
\newcommand{\Mpc}{\rm\,Mpc}
\newcommand{\sqdegr}{\raisebox{0.65ex}{\tiny\fbox{$ $}}\,$^{\circ}$}
\newcommand{\kmsMpc}{\rm\,km\,s^{-1}\,Mpc^{-1}}
\newcommand{\fratio}{\rm\,[f(8)/\rm\,f(3.6)]}
\newcommand{\fratioz}{\rm\,^{z}[f(8)/\rm\,f(3.6)]}
\newcommand{\fratiok}{\rm\,^{0.4}[f(8)/\rm\,f(3.6)]}

\begin{document}

\title{Unveiling the Important Role of Groups in the Evolution of Massive Galaxies: Insights from an Infrared Passive Sequence at Intermediate Redshift}

\author{D.~J.~Wilman$^1$,~D.~Pierini$^1$,~K.~Tyler$^2$,~S.~L.~McGee$^3$,~A.~Oemler~Jr$^4$,~S.~L.~Morris$^5$,~M.~L.~Balogh$^3$,
~R.~G.~Bower$^5$,~J.~S.~Mulchaey$^4$}
\affil{$^1$Max-Planck-Institut f\"ur extraterrestrische Physik, Giessenbachstra\ss e, D-85748 Garching, Germany.\\
$^2$Steward Observatory, University of Arizona, 933 North Cherry Avenue, Tucson, AZ 85721, U.S.A.\\
$^3$Department of Physics and Astronomy, University of Waterloo, Waterloo, Ontario, N2L 3G1, Canada.\\
$^4$Observatories of the Carnegie Institution, 813 Santa Barbara Street, Pasadena, California, U.S.A.\\
$^5$Physics Department, University of Durham, South Road, Durham DH1 3LE, U.K.\\
}

\begin{abstract}
The most massive galaxies in the Universe are also the oldest. 
To overturn this apparent contradiction with hierarchical growth models, we focus on the group-scale haloes which host most of these galaxies. 
Our $z\sim0.4$ group sample is selected in redshift space from the CNOC2 redshift survey. 
A stellar mass selected $\Mstellar\gtrsim 2\times10^{10}\Msol$ sample is constructed using IRAC observations. 
A sensitive Mid InfraRed (MIR) IRAC colour is used to isolate passive galaxies. 
It produces a bimodal distribution, in which 
passive galaxies (highlighted by morphological early-types) define a tight MIR colour sequence 
(Infrared Passive Sequence, IPS). 
This is due to stellar atmospheric emission from old stellar populations. 
Significantly offset from the IPS are galaxies where reemission by dust boosts emission at $\lambdaobs$=8\micron. 
We term them InfraRed-Excess galaxies whether star formation and/or AGN activity are present. 
They include all known morphological late-types. 
Comparison with EW[OII] shows that MIR colour is highly sensitive to low levels of activity, 
and allows us to separate dusty-active from passive galaxies at high stellar mass. 
The fraction of InfraRed Excess galaxies, $\fire$ drops with $\Mstellar$, such that $\fire=0.5$ at a ``crossover mass'' of $\Mcr\sim 1.3\times10^{11}\Msol$. 
Within our optically-defined group sample there is a strong and consistent deficit in $\fire$ at all masses, 
but most clearly at $\Mstellar\gtrsim 10^{11}\Msol$. 
Suppression of star formation must mainly occur in groups. 
In particular, the observed trend of $\fire$ with $\Mstellar$ can be explained if suppression of 
$\Mstellar\gtrsim 10^{11}\Msol$ galaxies occurs primarily in the group environment. 
This is confirmed using a mock galaxy catalogue derived from the Millenium Simulation. 
In this way, the mass-dependent evolution in $\fire$ (\emph{downsizing}) can be driven 
solely by structure growth in the Universe, as more galaxies are accreted into group-sized haloes with cosmic time. 
\end{abstract}

\keywords{galaxies: statistics --- galaxies: evolution --- infrared: galaxies --- galaxies: photometry --- galaxies: high-redshift --- galaxies: clusters: general}

\section{Introduction}\label{sec:intro}

Perhaps the biggest puzzle of galaxy formation in recent decades has been the early and rapid formation 
of the most massive galaxies \citep{Bender98,Cimatti04}. 
This apparently supported the ``monolithic collapse'' model \citep{Eggen62}, 
in contradiction to the standard picture of a Cold Dark Matter (CDM) Universe, 
in which galaxies grow hierarchically through mergers, along with the dark matter haloes in which they are 
embedded \citep[e.g.][]{Benson02}. 
Observationally, the picture has been embellished by recent galaxy surveys from $z\sim 0$ to $z \sim 1.5$. 
Going to higher galaxy mass and to denser environments, galaxies are older 
\citep{Kauffmann03,Panter03,Poggianti04,Thomas05,Salim05} 
and more likely to have experienced a shorter star forming life which is now truncated 
\citep{Yamada05,Baldry06,Pannella06,Haines07,Bundy06,Hopkins07}. 
The term \emph{downsizing} \citep{Cowie99} is used to describe the earlier and less extended star formation histories in 
more massive galaxies. 

Recent attempts to explain downsizing trends have examined the ways in which 
gas heating and cooling depend upon mass and redshift \citep[e.g.][]{Dekel06}, 
and invoked new heating methods such as feedback from Active Galactic Nuclei \citep[AGN,][]{Bower06,Croton06}. 
Regardless of the physics, a hierarchical Universe can itself supply a natural explanation for dependence of galaxy 
properties upon stellar mass and upon environment at the same time \citep{deLucia06}. 
As more and more massive galaxies are encorporated into more and more massive haloes, the probability 
of star formation being truncated is increased. 
Even in the populous loose group environment the properties of galaxies depend strongly upon environment 
\citep[e.g.][]{Postman84,Zabludoff98,Lewis02,Gomez03,Balogh03,Wilman05,Weinmann06,Gerke07}. 
Thus, evolution in groups can play an important role in driving evolutionary trends. 

The fraction of passive galaxies increases steadily and monotonically with increasing stellar mass. 
It is therefore convenient to quantify downsizing trends by identifying a stellar mass at which the fraction 
of passive (or red/early-type) galaxies is defined to be exactly 50$\%$. 
This has no physical significance, and so we prefer to call this the ``\emph{crossover mass}'' ($\Mcr$) 
rather than the ``\emph{transition mass}'' \citep[e.g.][]{Bundy06}. 
$\Mcr$ decreases strongly with increasing environmental density \citep{Baldry06,Bundy06,Haines07}, 
with a maximum $\Mcr \sim 2\times10^{10}\Msol$ in the \emph{lowest density environments} 
where the $z\sim0$ red galaxy fraction is 50$\%$ \citep{Baldry06}. 
The evolution of $\Mcr$ for the global population, 
indicates $\Mcr \sim 2.5-6\times10^{10}\Msol$ at 
$z \sim 0.4$ and $\Mcr \sim 7-10\times10^{10}\Msol$ at $z \sim 1$ \citep{Hopkins07}. 

These studies, mostly at rest-frame UV to optical wavelengths, are hindered by the
strong correlation between galaxy mass and dust attenuation in star-forming galaxies 
\citep{Giovanelli95,Wang96,Masters03,Brinchmann04,Weiner07}.
This correlation may arise directly from the variations of star
formation history with mass \citep{Calura07} and can lead to confusion
between truly passive and dusty star forming galaxies.  
In some cases light from the youngest stars can be totally removed from the line of sight, 
not only because their SED peaks at shorter wavelengths where dust attenuation is most effective \citep[e.g.][]{Pierini04}, 
but also because the HII regions in which they are embedded can be particularly dusty \citep{Duc02,Tuffs04}. 
This degeneracy of passive versus dusty star forming galaxies is well known at high
redshifts, where a selection of optically red galaxies (Extremely Red
Object/Galaxy -ERO/ERG or Distant Red Galaxy - DRG) is more and more
likely to select dusty starbursts as one moves to even higher redshift
(from $\sim 50\%$ at $z\sim1$ \citep{Cimatti02,Lotz06} to $\gtrsim 90\%$
at $z \sim 2-3$ \citep{Papovich06}, with an increasing contribution
from increasingly active infrared-bright galaxies 
\citep{Caputi06,Daddi07}). 

Even in the local Universe the optical red sequence \citep[e.g.][]{Bower92} 
is contaminated with star-forming
galaxies and some AGN. 
These ``interlopers'' can be identified using optical or UV colour-colour diagrams 
or high signal to noise emission line measurements \citep[e.g.][]{Wolf05,Haines07}. 
However the dust emission really stands out in the infrared where the
absorbed energy (mainly non-ionizing UV photons) is re-emitted
\citep[][and references therein]{Popescu00}. 
$31\%$ of red-sequence galaxies (by number density) are 24\micron-bright, 
corresponding to $\sim17\%$ star-forming galaxies and $\sim14\%$ AGN-host galaxies \citep{Davoodi06}. 

At shorter wavelengths ($\sim$3-12\micron) strong emission features exist in the presence 
of star formation \citep{Phillips84,Roche91}. 
However they are usually absent in the more 
highly ionized environments typical of AGN \citep{Aitken82,Roche84,Roche91}. 
These have been identified as PAH (Polycyclic-Aromatic-Hydrocarbon) features 
\citep[most obviously characterized by the strong, broad emission features at 6.2\micron\, 
7.7\micron\, 8.6\micron\ and 11.3\micron,][]{Leger84, Desert90}, and loosely trace star 
formation rates on the scale of a galaxy \citep{Roussel01}. 
Spectral diagnostics have been developed using the strengths of these features to distinguish 
star formation from nuclear activity \citep{Genzel98, Laurent00}. 
PAH carriers are apparently ubiquitous in high metallicity star-forming galaxies 
\citep[e.g.][]{Boselli98,Genzel00,Draine07}, although in low metallicity 
galaxies they may be absent \citep{Engelbracht05,Draine07}.
Calibration of physical quantities, such as star formation rate, from
PAH bands is complicated by the relative importance of circumstellar dust, 
and diffuse dust heated by the general radiation field or collisional impacts. 
These components may be significant 
for galaxies with low to moderate star formation rates and dust 
contents \citep[see e.g.][]{Boselli04,PerezGonzalez06,Calzetti07}.

To keep things simple, one can construct Mid-InfraRed (MIR) colours as opposed to calibrating star formation rates. 
The wealth of information in this spectral range prompted the use of MIR colours 
to trace star formation, dust temperature and nuclear activity \citep[e.g.][]{Dale00,Boselli03,Fisher06}. 
Normalizing the flux from a dust-sensitive band ($\lambdarest \gtrsim 6$\micron)
by a stellar mass-sensitive band (tracing old stellar populations, e.g. at 3.6\micron) 
produces a MIR colour highly sensitive to the level of dust emission per unit stellar mass. 
This is especially well correlated with morphology, tracing the warm dust emission from star-forming 
spiral arms \citep{Pahre04}. 
Indeed, in the local Universe, the MIR colour can effectively predict
morphology. 
\citet{Li07} find 88$\%$ early-types below 
$\rm \nu L_{\nu}(8)/\nu L_{\nu}(3.6) = 0.303$ and 83$\%$ late-types
above this division
\footnote{Includes assumed stellar contribution to $\rm \nu L_{\nu}$(8) of 0.232$\rm \nu L_{\nu}$(3.6).}. 
The combined sample exhibits a bimodal distribution in MIR colour. 
\citet{Johnson07} also see this bimodality in MIR colours for galaxies in Hickson Compact Groups, 
and show that the fraction of MIR-passive galaxies is much higher in the more evolved (HI-poor and X-ray bright) groups. 
These results demonstrate the power of IRAC colours to separate truly passive galaxies from star-forming or 
AGN-host galaxies 
without the dust-driven degeneracies of optical and UV studies. 

Our goal is to isolate truly passive galaxies and thus trace the
build-up of the increasingly important population of passively evolving
galaxies in the Universe, as a function of environment. 
In particular, we wish to study massive ($\Mstellar \gtrsim 2 \times 10^{10}\Msol$) 
galaxies at $z\sim0.4$, where the contributions of both passive and dusty star-forming galaxies 
are important, but can be confused with each other without due attention. 
Separating these populations using the MIR colour will alleviate this problem. 
Strong evolution in the dust and star formation properties of normal galaxies 
over the range $0<z\lesssim 0.7$ is expected: 
At $z\gtrsim 0.7$ spirals dominate the Luminous InfraRed Galaxy (LIRG) population 
 \citep{Bell05,RowanRobinson05,Zheng07}, 
whilst today galaxies of equivalent infrared luminosities are much rarer, and 
are typically interacting systems and mergers. 
Our galaxy sample, selected from the CNOC2 redshift survey \citep{Yee00}, 
probes intermediate redshifts ($z \sim 0.4$). 
This redshift corresponds to a look-back time of $\sim 3-4\Gyr$, when 
the volume-averaged star formation rate in the Universe was 
significantly higher than it is today \citep{Lilly96,Madau98,Hopkins04}. 

In Section~\ref{sec:sample} our sample is introduced, including archive IRAC data in our 
fields. 
From this stellar masses and MIR colours are derived, a mass-selected sample is constructed. 
Selection effects will be discussed in detail. 
Section~\ref{sec:colmass} presents the MIR colours of $z\sim0.4$ galaxies as a function of their 
stellar masses and morphologies. 
A tight \emph{Infrared Passive 
Sequence} is identified in the MIR colour-mass plane as a tool to select passive galaxies. 
Section~\ref{sec:c41oii} examines the correlation of MIR colour with the EW[OII]$\lambda3727$ emission line 
diagnostic, testing their relative sensitivity. 
In Section~\ref{sec:fracIR} we compute the fraction of massive galaxies with 
InfraRed Excess (IRE galaxies) as a function of stellar mass. 
The role of the group environment is discussed in Section~\ref{sec:environment},
where we examine galaxies in our optical group catalogue \citep{Carlberg01}.
Finally, our results are discussed in the context of our current understanding of 
galaxy evolution and the group environment in Section~\ref{sec:discussion}. 
In particular, by applying a simple model to a mock galaxy catalogue, 
we focus on the possibility that downsizing is intimately related to, 
if not entirely caused by, the growth of structure with time. 
Our conclusions are presented in Section~\ref{sec:concl}.
Throughout this paper a $\rm \Lambda CDM$ cosmology with 
$\rm \Omega_m =$ 0.3, $\rm \Omega_\Lambda =$ 0.7 and 
$\rm H_0$ = 100$h\kmsMpc$ with $h$=0.75 is assumed. 

\section{Sample}\label{sec:sample}

The objective of this paper is to study MIR activity in galaxies as a function of 
their stellar mass and environment. 
This is facilitated by the construction of a mass-selected sample in the redshift 
range $0.3\leq z \leq 0.48$, described in this section.

\subsection{CNOC2 Survey and Groups}\label{sec:CNOC2}

Our sample is based upon the CNOC2 redshift survey which is magnitude
limited down to $\Rc\sim23.2$ in photometry and totals $\sim$1.5\sqdegr\ within four separate patches on the sky. 
Spectroscopic redshifts exist for a large and unbiased sample of $\Rc \lesssim 21.5$ galaxies \citep{Yee00}. 
Within the fields of 20 kinematically selected groups from the catalogue of 
\citet{Carlberg01}, the original $\UBVRI$ photometry and spectroscopy have been supplemented with targetted
spectroscopy to $\Rc=22$ using the Magellan 6.5m telescope \citep{Wilman05}, and F775W-band HST ACS imaging. 
Galaxies with spectroscopic redshifts $z>0.3$ and ACS coverage have been visually morphologically classified by one of 
the authors (Augustus Oemler Jr, hereafter AO). 

The probability that a galaxy has a redshift is a function of its $\Rc$-band magnitude, 
and is unbiased to $\Rc=21.5$. 
To study the statistical properties of galaxies as a function of their stellar mass (and not $\Rc$-band magnitude), we must statistically correct for this selection function and magnitude limit. 
This is achieved by weighting galaxies by the inverse probability that they appear in the $\Rc\leq 21.5$ spectroscopic sample. 
There are two components to this weight:
\begin{itemize}
\item{A weight $\rm W_{\Rc}$ to account for the fraction of galaxies which are targetted spectroscopically (with measured redshifts) as a function of $\Rc$-band magnitude. Since below $\Rc=21.5$ the success of redshift determination may be biased in favour of emission line galaxies, $\rm W_{\Rc}$ is set to 0 for $\Rc>21.5$.}
\item{A weight $\rm W_z$ to account for the fraction of sample volume within which a galaxy would fall out of our $\Rc\leq21.5$ sample (effectively a $\rm V_{max}$ correction). For a $0.3\leq z \leq 0.48$ sample, this means that for each $\Rc\leq 21.5$ galaxy in the sample, $\rm W_z =$ $\rm V$($\rm z=0.3$ to $\rm z=0.48$)$\rm /V$($\rm z=0.3$ to $\rm z=z_{lim}$) where V is the cosmological volume in our chosen cosmology, and $\rm z_{lim}$ is the redshift at which the galaxy would have a magnitude $\Rc=21.5$.}
\end{itemize}
Then the final weight for each galaxy is simply: $\rm W = W_{\Rc} \times W_z$.

\subsection{IRAC Photometry}\label{sec:irac}

The \emph{Spitzer Space Telescope} GTO program 64 (PI Fazio) included an IRAC \citep[Infrared Array Camera,][]{IRAC} 
scan-map survey of $\sim$20\arcmin$\times$30\arcmin\ regions in the centre of three of the four CNOC2 patches. 
The IRAC focal plane geometry translates each mapping into one mapping of bands 1 (3.6\micron) and 3 (5.8\micron) and 
another mapping of bands 2 (4.5\micron) and 4 (8.0\micron), offset by 6.7\arcmin. 

The astrometrically and photometrically calibrated post-BCD (Post -- Basic Calibrated Data) products 
were retrieved from the Spitzer archive. The pipeline version used was S11.4.0. 
The processed images are clean, with no image artifacts, and provide stable source flux measurements and accurate sub-arcsecond astrometry. 
Apart from at the map edges, all sources receive $\sim$500~s imaging time in all four bands 
(100~s frame-time at five dithered positions). 
The complete four band IRAC coverage has significant ($\geq 50\%$) overlap with six ACS fields, 
where galaxies have also been morphologically classified.  
Shallow MIPS 24\micron\ imaging is also available, but with lower sensitivity and poorer 
diffraction-limited resolution than the IRAC bands. 

For all sources detected in the 3.6\micron\ frame, colours are measured
using the {\it SExtractor} package \citep{Bertin96} in {\it double-image} mode. 
They are derived from fluxes measured inside 3\arcsec\ radius apertures centred 
at the position of all 3.6\micron\ detected objects, in all IRAC bands. 
Detection requires at least four adjacent pixels which are at least 1.5-$\sigma$ above the background level at 3.6\micron. 
Aperture corrections were 
computed to correct for light lost from the aperture due to the spatial extent 
of the PRF (Point Response Function)\footnote{Measured by combining
  larger aperture measurements for isolated stars with IRAC Data
  Handbook values. Corrections are 1.184, 1.180, 1.249 and 1.404 for
  IRAC bands 1-4 respectively.}. 
Corrected measurements are used to compute flux ratios (colours) within the aperture. 
The {\it SExtractor} FLUX\_AUTO Kron-like measurement provides more
stable total flux measurements, and thus is used to compute absolute
galaxy luminosity at 3.6\micron. 
We follow the practice of 
SWIRE\footnote{swire.ipac.caltech.edu/swire/astronomers/publications}
and compute the flux error from the measurement image itself, weighted
by inverse variance using the coverage map. 

To match the 3.6\micron\ selected catalogue to the CNOC2 redshift catalogue, 
we apply the {\it tmatch} tool in {\it iraf}
\footnote{IRAF is distributed by the National Optical Astronomy Observatories,
    which are operated by the Association of Universities for Research
    in Astronomy, Inc., under cooperative agreement with the National
    Science Foundation}. 
None of the galaxies in our spectroscopic sample are lost from the IRAC sample due to lack of sensitivity except at the edge of the IRAC field. 
Visual inspection shows that CNOC2 galaxies with no match either lie outside the IRAC zone of full coverage, can be identified with sources improperly deblended by {\it SExtractor}, or lie in regions of particularly poor CNOC2 astrometry. 
The sample is then limited to those objects with unambiguous matches, 
with no apparent bias relative to the $\sim 15\%$ of CNOC2 galaxies with IRAC coverage which are lost at this stage. 
However, to maintain statistical accuracy the weight $\rm W_{\Rc}$ is modified to account for galaxies without spectroscopy or 3.6\micron\ matches, as a function of $\Rc$-band magnitude.
At 3.6\micron, the 2-$\sigma$ depth is $\sim$1$\mu$Jy 
(for a passive galaxy at z$=$0.4, this corresponds to a stellar mass of $\sim9\times10^8\Msol$). 

The goal is to construct a $0.3\leq z \leq 0.48$ mass-selected sample for which representative spectroscopy is available, and reaching required depth at 8\micron\ to distinguish passive galaxies from those with excess emission at 8\micron\ (2-$\sigma$ depth of $\sim$10$\mu$Jy). 
Effectively, these criteria set the mass limit to $2\times10^{10}\Msol$, well above the 3.6\micron\ limit (\S~\ref{sec:samplesel} for more details).

\subsection{Deriving Stellar Masses and MIR Colours}\label{sec:irderiv}

\subsubsection{Stellar Mass}

Near InfraRed (NIR) light is a useful tracer of stellar mass because of the weak dependence on 
star formation history \citep{Aaronson79,Rix93,Bell01}. 
3.6\micron\ flux has been converted into luminosity and thence to stellar mass 
following \citet{Balogh07}. 
To give a brief overview of this procedure: 
Observed-frame AB luminosities are computed for the assumed cosmology, and an empirical k-correction is applied 
to correct to a consistent rest-frame of a $z=0.4$ galaxy. 
In this paper, we will call this $^{0.4}$[3.6]. 
Secondary corrections are then applied to move into the K-band rest-frame. 
These corrections are calibrated for objects with both [3.6] and K-band photometry (not used in this paper). 
The mass to light ratio of each galaxy in K-band ($\Mstellar$$\rm /L_K$) is estimated using a combination of two simple \citet{BC03} models with a \citet{Chabrier03} IMF. 
For blue galaxies ($\rm B - V < 0.4$), a constant star formation model with dust extinction $\tau_v = $1 mag gives $\Mstellar$$\rm /L_K$ = 0.2. 
For red galaxies ($\rm B - V > 1.0$), a 11.7-Gyr old, dust free single stellar population model gives $\Mstellar$$\rm /L_K$ = 0.7. For intermediate colours a linear interpolation between these values is used. For more details, please refer to \citet{Balogh07}. 
This means that the full range of galaxy types only exhibit a range in $\Mstellar\rm /L_K$ (and thus also 
$\Mstellar\rm /L_{^{0.4}[3.6]}$) of 3.5. 
Therefore our results are relatively insensitive to whether we select in mass or $^{0.4}$[3.6] luminosity. 
We prefer to select in mass as it is the more physical quantity and, thus, allows a comparison with the literature. 

\subsubsection{MIR colour}
\begin{figure*}
  \epsscale{0.80}
  \plotone{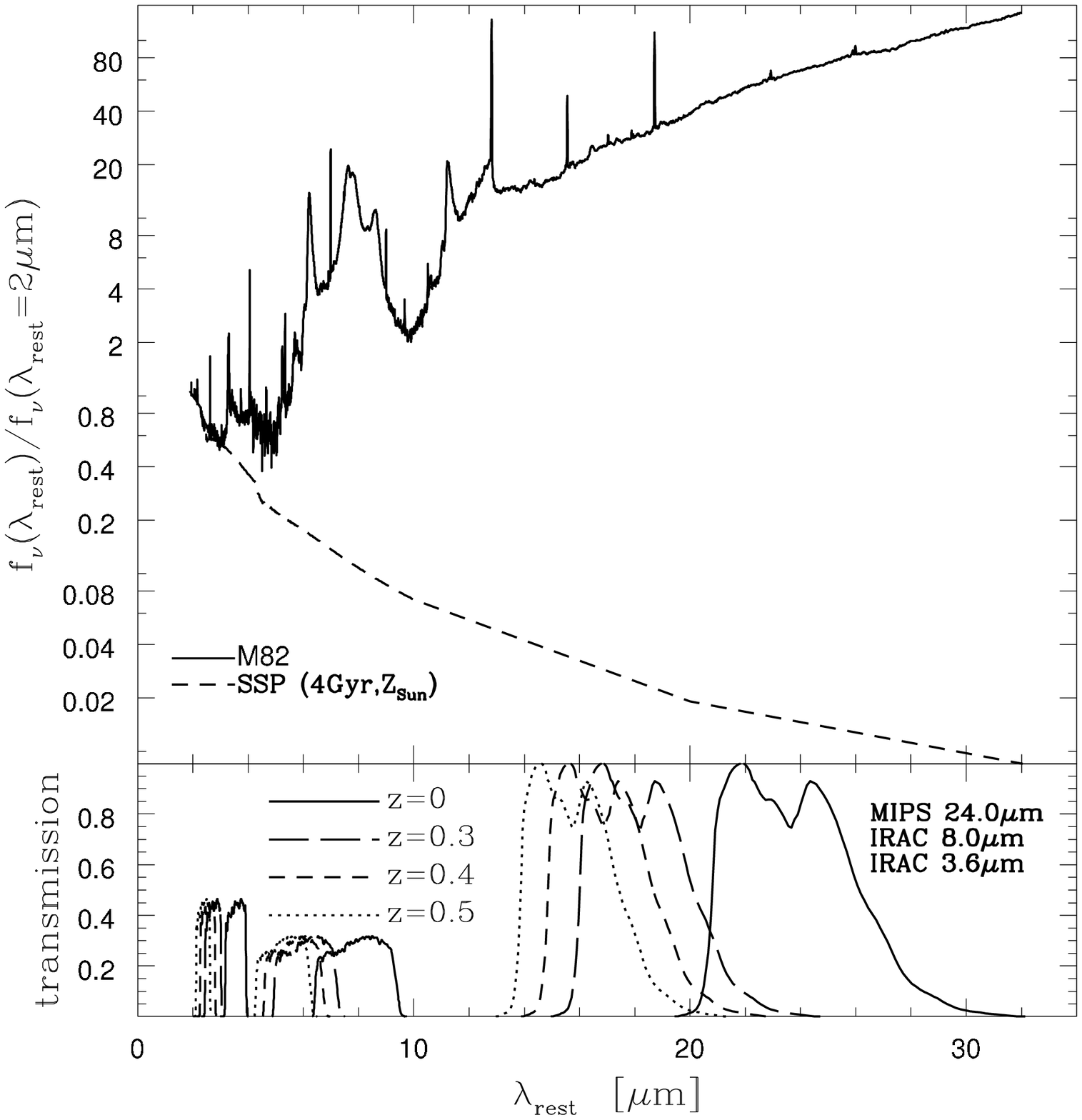}
  \caption{Our two MIR spectral templates are illustrated in the {\bf
    top panel}. A 4$\Gyr$ old Simple Stellar Population traces pure stellar 
    atmospheric emission from old stars (SSP, dashed line). 
    This is generated using \emph{PEGASE2} \citep{Fioc97PEGASE2}, 
    with solar metallicity, and a \citet{Salpeter55} IMF with upper mass limit of 120$\Msol$. 
    The other template is the ISO-SWS spectrum of the dusty star
    forming galaxy M82 \citep[solid line, ][]{FSchreiber01}. 
    The data is limited to $\rm \lambda \geq 2.4$\micron, and so we 
    extrapolate down to 1.9\micron\
    by combining it with the spectrum obtained from a 
    composite stellar population model (constant star formation rate,
    evolving metallicity, 1$\Gyr$ since start of star formation, Salpeter IMF
    with upper mass limit of 120$\Msol$).
    The two templates are normalized to match at $\rm \lambda = 2.2$\micron. 
    Beyond $\sim 3$\micron,
    the spectrum of M82 is dominated by dust emission in the form of PAH features
    ($\lesssim 12$\micron) and a warm dust emission continuum ($\gtrsim
    12$\micron).  The {\bf bottom panel} shows the effective
    transmission functions of Spitzer IRAC bands at 3.6\micron\ and 
    8\micron, and the MIPS 24\micron\ band in the
    \emph{rest-frames} of galaxies at $z=0$ (solid), and $z=0.3$ (long-dashed),$0.4$ (short-dashed) and
    $0.5$ (dotted), corresponding to the redshift range of our galaxies. 
    Whilst 3.6\micron\ traces stellar emission, 8\micron\ and 24\micron\ trace PAH and warm
    dust emission respectively at $z\sim 0.4$.}
  \label{figure:templates}
\end{figure*}

The two template spectra in the top panel of Figure~\ref{figure:templates} demonstrate the diversity and wealth of spectral 
information at MIR wavelengths. 
A passive, dust-free galaxy is likely to exhibit a spectrum similar to the SSP template (dashed line). 
With no dust emission, the infrared emission originates solely from
photospheres of the cold stellar population, primarily giant M-stars. 
Conversely, the spectrum of a dusty star forming galaxy should look more like that of M82 (solid line). 
Beyond $\lambdarest \sim 3-4$\micron, a typical dust emission spectrum 
with characteristic PAH features comes to dominate the M82 template.

The lower panel of Figure~\ref{figure:templates} shows the transmission function of the IRAC bands at 3.6\micron\ and 8\micron, 
and the MIPS \citep[Multi-Band Imaging Photometer for Spitzer,][]{MIPS} 
24\micron\ band. 
Each transmission function is transformed to indicate the rest-frame wavelength domain probed 
for a galaxy located at $z=0.3$,$0.4$ or $0.5$. 
It is clear that excess (enhanced beyond pure stellar photospheric) emission can be revealed at both 8\micron\ and 24\micron\ 
for galaxies within the redshift range of interest. 
For $0.05 \lesssim z \lesssim 0.48$ galaxies, the 8\micron\ band captures the strong 6.2\micron\ PAH feature, 
tracing star formation \citep[e.g.][]{Peeters04,FSchreiber04}. 
Crucially, stellar atmospheric emission drops off significantly from 8\micron\ to 24\micron: 
Thus detection of passive galaxies, and thence accurate separation of passive from star-forming or AGN-host galaxies, 
requires much shallower exposures at 8\micron\ than 24\micron. 
A colour constructed from 8\micron\ and 3.6\micron\ has the leverage on the galaxy spectrum to make this distinction. 
In this paper we focus on IRAC data for these reasons. 

For the sake of simplicity, we compute k-corrected MIR colours of our galaxies
using a linear combination of the two simple templates in Figure~\ref{figure:templates} 
\citep[similar to the approach of ][]{Huang07}. 
We define our colour to be the flux-ratio $\fratio$
\footnote{for a more traditional colour, simply compute [3.6]-[8] = +2.5log$_{10}$($\fratio$)}. 
For each galaxy, the quantity $\eta$ is measured: 
This is the required fractional contribution from the M82 template to match the {\it observed-frame} $\fratioz$
\footnote{The prefix superscipt z is used to indicate the band-pass \citep{Blanton07kcor}; i.e. the redshift of a galaxy for which the k-correction is defined to be zero.}. 
Our k-corrections are based on the assumption that $\eta$ (approximately {\it dust to stellar flux ratio}), 
would be the same regardless of the redshift of the galaxy (and thus the part of the spectrum sampled). 
$\eta$ is computed as:

\begin{equation}
  \label{eq:eta}
  \eta = \frac{\fratioz-\fratioz(SSP)}{\fratioz(M82)-\fratioz(SSP)}
\end{equation}

To minimize the k-corrections $\fratioz$ is transformed to the equivalent colour for a z=0.4 galaxy, 
and limit ourselves to galaxies in the narrow redshift range $0.3\leq z \leq0.48$, 
an epoch $\gtrsim 3\Gyr$ ago. 
The k-corrected $\fratio$ colour is computed as:

\begin{equation}
  \fratiok = \eta \times \fratiok(M82) + (1 - \eta)\times \fratiok(SSP)
  \label{eq:c41z}
\end{equation}

A pure SSP k-correction ($<0.07\fratioz$) increases by $\lesssim 10\%$ going from ages of 8$\Gyr$ to 1$\Gyr$, and $10-20\%$ going
from solar to $3\times$ solar metallicity. 
For $\eta=1$ (M82-like), $\fratiok$=2.17 ($\sim 7$ times the stellar photospheric continuum emission 
within the IRAC 8\micron\ band). 
Whilst k-corrections of high $\eta$ galaxies are more uncertain due to the assumed M82 template, 
k-corrections remain relatively small within our redshift range ($\lesssim 50\%$ for $\eta=1$). 
In this paper we will only separate dusty galaxies emitting excess light at infrared wavelengths from
passive galaxies, and will not draw strong conclusions from precise 
MIR colours of dusty galaxies which are sensitive to the applied k-correction. 
To this extent, we believe that we are justified in applying these rough k-corrections.

\subsection{A Mass-Selected Sample}\label{sec:samplesel}

In this paper, our analysis is limited to the subsample of galaxies with known
redshifts $0.3\leq z \leq 0.48$, and with full IRAC coverage.  Of
these, the 333 galaxies with stellar masses $\Mstellar>2\times10^{10} \Msol$ are selected, 
corresponding to a luminosity of $^{0.4}$[3.6]=-20.89 for passively evolving galaxies, 
or a flux of $\rm f(3.6\micron)>20 \mu$Jy, well above the detection limit. 

This stellar mass limit is actually defined by the depth of the
$8\micron$ data, since we require this colour information for our
analysis.  In our mass-limited sample, 313 of the 333 galaxies are
detected ($2 \sigma$) at $8\micron$, and the remainder have $>1.8\sigma$ 
upper limits that are sufficient to identify them as passively evolving
galaxies, characterised by colours $^{0.4}\fratio<0.5$ 
(see \S~\ref{sec:colmass} for a justification of this definition). 

The requirement that all galaxies have a spectroscopic redshift introduces sensitivity to the optical ($\Rc$) selection function. 
To account for selection bias, the spectroscopic selection weight $\rm W$ is applied (\S\ref{sec:CNOC2}). 
This weight includes the factor $\rm W_z$ which applies a volume correction to correct for galaxies which fall out of the $\Rc\leq21.5$ sample at $\rm z_{lim}\leq0.48$. 
Within a $\Mstellar \geq 2\times10^{10} \Msol$ sample, the lowest mass and highest redshift galaxies with high mass to light ratios will not make this magnitude cut. 
Nonetheless, the weighted sample is still representative: 
A passively evolving $\Mstellar=2\times10^{10} \Msol$ galaxy can be sampled with $\Rc \leq 21.5$ up to a redshift of $\rm z_{lim} \sim 0.38$ ($\sim36\%$ of the volume sampled) and receives a weight $\rm W_z = 1./0.36 = 2.78$. 
There do exist four $\Rc \leq 21.5$ galaxies with unusually high mass to optical light ratios. 
These can more easily be lost from the sample due to their extremely red optical-NIR colours. 
Such red colours can exist where there is extreme dust extinction, or a population of AGB stars boosting NIR luminosity. 
However, they are too few to significantly influence our statistics. 
Our sample of 333 galaxies includes eleven $\Mstellar>2\times10^{10} \Msol$ galaxies with $\Rc>21.5$. 
These are assigned zero weight, but it is worth noting that the IRE fraction of these optically faint galaxies is comparable to that of other galaxies. 
In practice, our results are not strongly sensitive to the statistical weights.

\section{Bimodal MIR Colours and a Passive Sequence}\label{sec:colmass}

\begin{figure*}
  \epsscale{0.80}
  \plotone{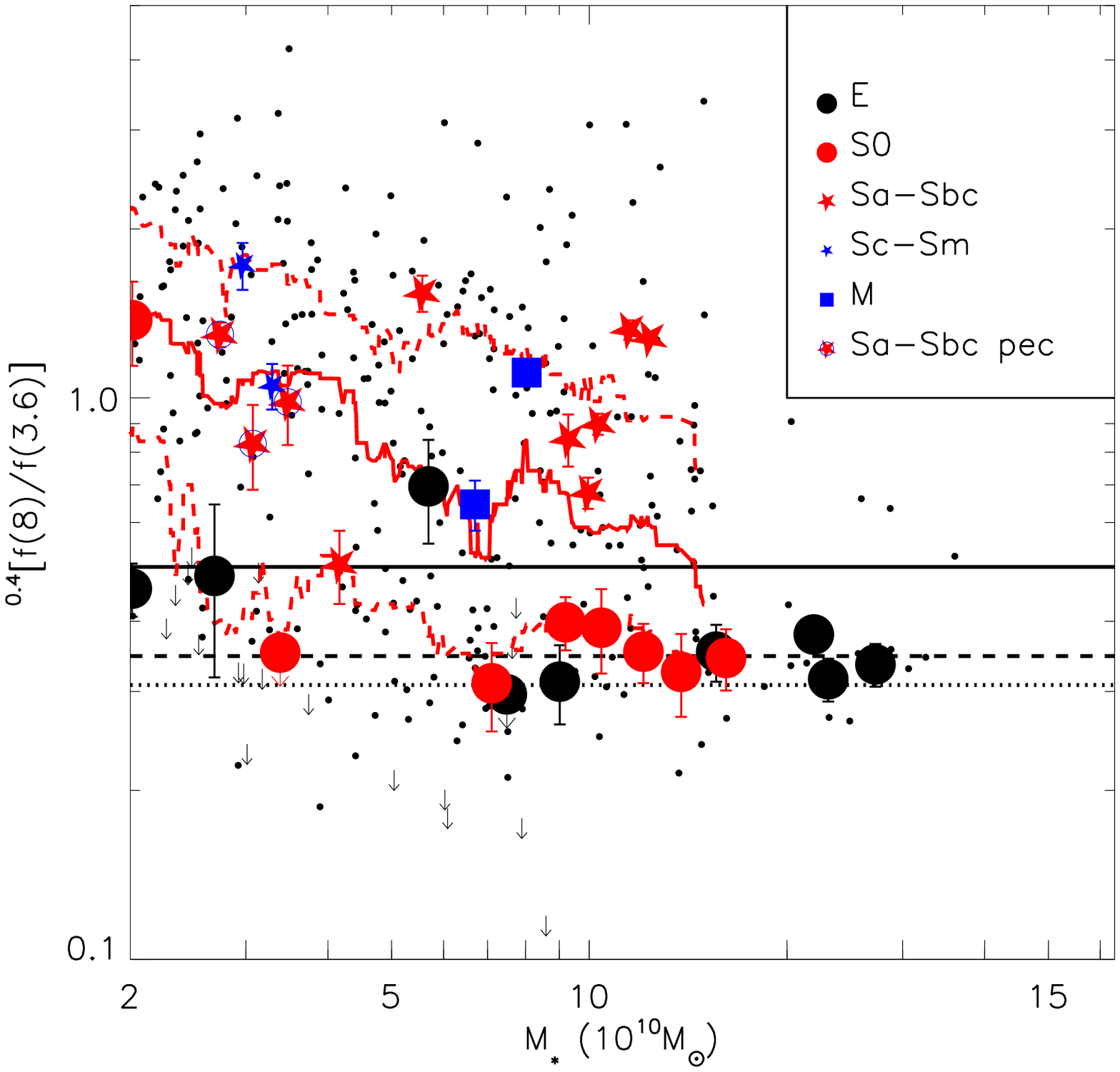}
  \caption{The $\fratiok$-$\Mstellar$ ``colour-mass''
    plane for our sample of 333 $\Mstellar\geq 2\times 10^{10}\Msol$, $0.3\leq z\leq 0.48$ galaxies. 
    For the few galaxies without 2$\sigma$ 8\micron\ detections, 
    2$\sigma$ upper limits on $\fratiok$ are indicated. 
    $25^{th}, 50^{th}$ and $75^{th}$ percentiles of the $\fratiok$ distribution are overplotted (red, solid and dashed lines). 
    These are unweighted, and computed for running bins of 50 galaxies in $\Mstellar$, 
    and undetected galaxies are considered to be located at their measured values as opposed to upper limit values. 
    Larger symbols are used for galaxies in the joint IRAC-ACS sample for which galaxies are 
    morphologically classified (see key). 
    For morphologically classified galaxies, measurement errors on $\fratiok$ are shown, which 
    are typical of the overall population. 
    Early-type galaxies cluster along a tight sequence at $\fratiok\sim0.35$, consistent with pure
    stellar photospheric emission (horizontal dashed line). 
    We call this the {\bf Infrared Passive Sequence (IPS)}, 
    and define InfraRed Excess (IRE) galaxies as those with  $\fratiok>0.5$ (horizontal solid line). 
    The SSP template is indicated by the horizontal dotted line ($\fratiok=0.308$).}
  \label{figure:c41mass}
\end{figure*}

In this section the distribution of the MIR colour $\fratiok$, and its dependence 
on galaxy stellar mass $\Mstellar$ is investigated. 
Figure~\ref{figure:c41mass} shows how the 333 galaxies in our 
$\Mstellar\geq 2\times 10^{10}\Msol$, $0.3\leq z\leq 0.48$ sample populate the $\fratiok$-$\Mstellar$ ``colour-mass'' plane, along with the $25^{th}, 50^{th}$ and $75^{th}$ percentiles of the distribution.  

The distribution shifts to lower $\fratiok$ with higher mass. 
Dust flux per unit stellar mass can be approximated by $\fratiok_{50\%}-0.308$ 
where the stellar atmospheric contribution to $\fratiok$ is estimated by the SSP 
template to be $\fratiok(SSP)=0.308$ (dotted line). 
The median relation between $\fratiok-0.308$ and stellar mass can be 
well fit by a linear relation in log space: 
$log_{10}(\fratiok_{50\%}-0.308) = 8.8\pm0.2 - (0.85\pm0.02) log_{10}(\Mstellar/\Msol)$. 
This relation is significantly shallower 
than would be expected for no relation between dust flux and stellar mass ($\fratiok-\fratiok(SSP)\sim \frac{1}{\Mstellar}$). 
However, this is sensitive to the exact value of $\fratiok(SSP)$ chosen. 

We also examine whether the strong correlation between MIR activity and morphology observed in the local 
Universe \citep{Pahre04,Li07} extends to our intermediate redshift sample. 
31 of the galaxies in our sample are within the ACS fields and have been morphologically classified by AO. 
These are indicated in Figure~\ref{figure:c41mass}. 

\begin{figure*}
  \epsscale{0.8}
  \plotone{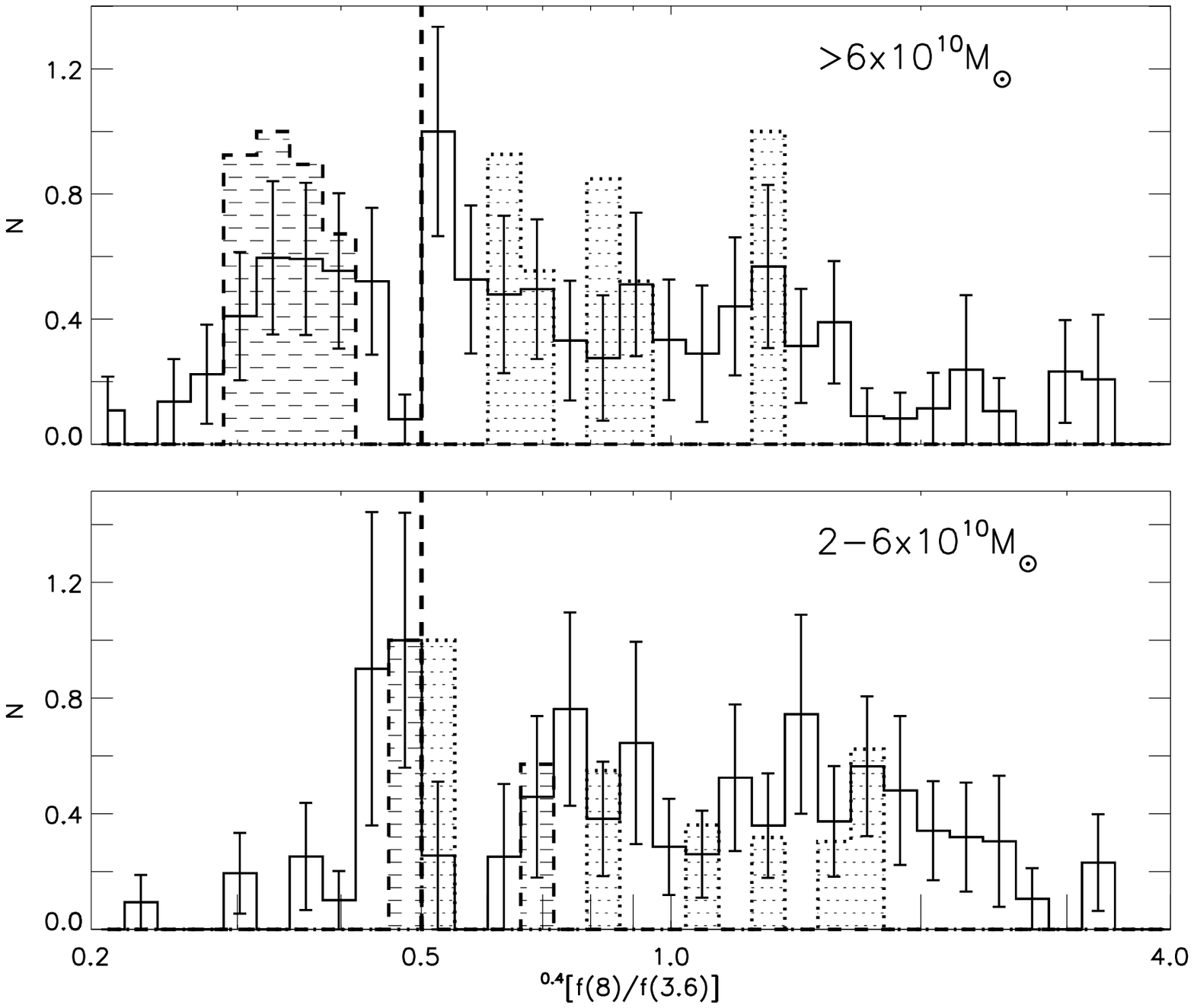}
  \caption{The weighted distribution of the MIR colour $\fratiok$ 
    for $0.3\leq z\leq 0.48$ galaxies with $\Mstellar>6\times 10^{10}\Msol$ {\bf (top)} and 
    $\Mstellar=2-6\times 10^{10}\Msol$ {\bf (bottom)}, 
    binned in log-space and with poissonian error-bars. 
    The few galaxies without 2$\sigma$ detections at 8\micron\ are excluded. 
    Since these are all constrained at the $\rm \geq 1.8\sigma$ level to 
    lie at $\fratiok<0.5$, their inclusion would not destroy the bimodality. 
    Overplotted are the distributions for morphologically classified early type (dash-filled) 
    and late type (dot-filled) galaxies, renormalized to give the same maximum value. 
    Our criteria for an InfraRed Excess galaxy ($\fratiok=0.5$) is indicated with the vertical dashed line, 
    and effectively separates the peak of the {\bf Infrared Passive Sequence (IPS)} ($\fratiok<0.5$) from the 
    InfraRed Excess (IRE) galaxy population.
}
  \label{figure:c41distr}
\end{figure*}

Morphologically classified early-type (Elliptical and S0) and late-type (mainly spiral) 
galaxies exhibit very different MIR colour distributions, as seen locally by \citet{Li07}. 
In fact, the overall distribution in $\fratiok$ is bimodal in nature, with the two peaks in MIR colour 
almost equivalent to these two morphological classes. 
This bimodality has been observed in other (local) optical-NIR selected samples \citep{Li07,Johnson07} 
whilst it is less obvious in shallow 8\micron-selected samples where early-types are more difficult to detect \citep{Huang07}. 
The bimodality of CNOC2 galaxies is demonstrated in Figure~\ref{figure:c41distr}, split into two bins of stellar mass. 
Within each mass range there exists a significant population of galaxies with $\fratiok<0.5$ and then a gap before the 
onset of a second population of galaxies emitting more strongly at 8\micron. 
Whilst the full distribution seems to shift to higher $\fratiok$ at lower mass, 
the value $\fratiok=0.5$ is within the gap across our mass range. 
Thus we adopt this value to divide the sample: Everything with $\fratiok>0.5$ we call an 
{\bf InfraRed-Excess (IRE)} galaxy. 
This division at 0.5 excludes galaxies from the passive population if they have $44\%$ more 8\micron\ 
flux than the early-type locus. 
Contribution from circumstellar dust \citep{Piovan03,Temi07} or silicate emission from mass-loss \citep{Bressan07} 
are unlikely to produce such a strong boost to the 8\micron\ flux.

The dash-filled histogram in Figure~\ref{figure:c41distr} indicates the $\fratiok$ distribution for morphologically 
early-type galaxies. 
The tight distribution of high mass early-types is mirrored in the narrow locus of almost constant colour 
in the colour-mass plane (Figure~\ref{figure:c41mass}). 
The 11 early-types (ignoring the single upper limit) with $\Mstellar>6\times10^{10}\Msol$ 
have mean colour and scatter $\fratiok\sim0.347\pm0.03$ (horizontal dashed line in Figure~\ref{figure:c41mass}). The scatter is consistent with the photometric errors. 
We call this the {\bf Infrared Passive Sequence (IPS)}, which is not far offset from the template SSP colour 
($\fratiok=0.308$, horizontal dotted line). 
The peak value and scatter of the IPS for the full population of $\Mstellar>6\times 10^{10}\Msol$ galaxies is $\fratiok\sim0.34\pm0.05$. 
The tight scatter of the IPS is reminiscent of the optical red sequence \citep[e.g.][]{Bower92}. 
Indeed the tightness of the IPS is due to the emission from 
stellar atmospheres of cold stars in passive galaxies, 
which is relatively little sensitivity to age or metallicity. 
Mid Infrared wavelengths are highly sensitive to dust
emission; thus the Infrared Passive Sequence does not suffer from
the degeneracy between dusty red galaxies and passive red galaxies as
by a red sequence defined at optical wavelengths. 

The second, broader peak at $\fratiok>0.5$ contains IRE galaxies with excess (non-stellar) emission at 8\micron. 
This excess emission traces some kind of activity, i.e. star formation or AGN. 
Figures~\ref{figure:c41mass} and~\ref{figure:c41distr} show that all known late-type galaxies 
(dot-filled histogram in Figure~\ref{figure:c41distr}) in the sample inhabit this peak. 
Two morphologically classified early-type galaxies also exhibit excess emission at 8\micron. 
Excess MIR emission in some early-type galaxies is known to exist, and correlates with low levels of star formation or 
AGN activity \citep[e.g.][]{Pahre04b,Bressan07}.  

\section{The $\fratiok$-EW[OII] plane}\label{sec:c41oii}

\begin{figure*}
  \epsscale{0.80}
  \plotone{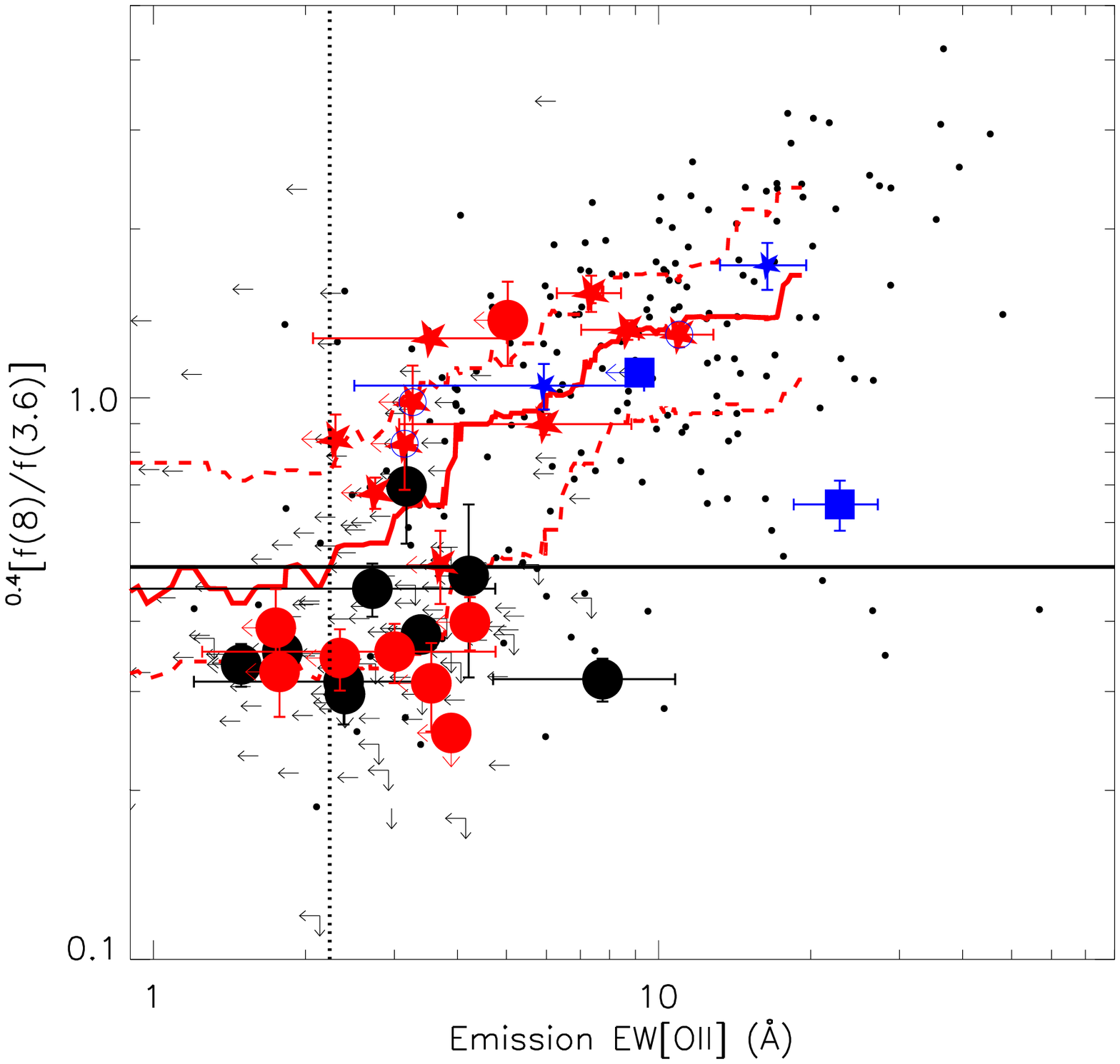}
  \caption{The $\fratiok$ - EW[OII] plane for 319 $\Mstellar \geq 2\times10^{10}\Msol$ and $0.3\leq z\leq 0.48$ galaxies. 
    $25^{th}, 50^{th}$ and $75^{th}$ percentiles of the $\fratiok$ distribution are overplotted, computed over running
    bins of 50 galaxies in EW[OII] (red, solid and dashed lines). 
    Undetected galaxies (at the 1$\sigma$ level in [OII] or the 2$\sigma$ level at 8\micron) are plotted at 
    their 1$\sigma$ EW[OII] and 2$\sigma$ $\fratiok$ upper limit levels (arrows), 
    whilst the percentiles are computed based upon \emph{measured} [OII] and 8\micron\ flux levels: on a statistical basis the 
    $50^{th}$ percentile should therefore be representative of the true median relation (both measured [OII] and 
    8\micron\ flux can be negative for non-detections). 
    Our empirical division ($\fratiok=0.5$) is overplotted (solid horizontal line). 
    There is a strong correlation between EW[OII] and
    $\fratiok$, inferring that for the average galaxy both diagnostics
    trace the same phenomenon (primarily star formation). 
    The median galaxy at our empirical division $\fratiok=0.5$ has
    EW[OII]$=2.1$\AA\ (dotted vertical line). 
    Larger symbols are used for morphologically classified galaxies as in Figure~\ref{figure:c41mass}, 
    with measurement errors typical of the overall population. 
  }
  \label{figure:c41oii}
\end{figure*}

It is useful to compare the MIR colour $\fratiok$ with an independent, 
well-studied indicator of activity (primarily star formation), the equivalent width of the [OII]$\lambda3727$\AA\ emission line. 
[OII] is most commonly used at z$\gtrsim$0.3 where the H$\alpha\lambda6563$\AA\ emission line is redshifted out of the optical window. 
By normalizing the emission flux by the continuum level, the sensitivity to flux calibration and dust absorption is reduced, producing a normalized quantity more easily comparable to $\fratiok$.

Figure~\ref{figure:c41oii} shows the location of our $\Mstellar \geq 2\times10^{10}\Msol$ and $0.3\leq z\leq 0.48$ galaxies 
(excluding 14 galaxies where EW[OII] could not be measured) in the $\fratiok$-EW[OII] plane, including $25^{th}$,$ 50^{th}$ and $75^{th}$ percentiles and morphologically classified galaxies overplotted with their errors. 

There is a strong correlation between EW[OII] and $\fratiok$. 
Both diagnostics trace galaxy activity, and primarily star formation, and so a strong correlation is expected. 
A value of EW[OII]$=2.1$\AA\ (dotted vertical line) corresponds to the median galaxy with 
$\fratiok=0.5$ (our empirical division in MIR colour, solid horizontal line), 
which is below the [OII] detection limit of most spectroscopic surveys. 
Indeed, we are only able to identify this value for a large statistical sample: The median error on EW[OII]
for individual galaxies in our sample is 2.4\AA. 

In \citet{Balogh07}, we divided our sample into passive and [OII]-strong galaxies at EW[OII]=10\AA. 
This is the equivalent of a division at $\fratiok=1.31$ for the median galaxy, selecting only
the tail of galaxies with the most ongoing activity. 
A division at EW[OII]=5\AA\ \citep[as used in][]{Wilman05} corresponds to $\fratiok=0.92$.
Therefore with a reasonable photometric depth at $\lambdarest \sim$6\micron, 
it is possible to pick up levels of activity not possible without extremely deep
optical spectroscopy. 
This inevitably leads to a higher fraction of galaxies defined as non-passive.
Many spiral galaxies (mainly of type Sa-Sbc) with $\fratiok>0.5$ have no detectable [OII] emission. 
Based upon a simple cut in EW[OII], these would be classified as passive spirals \citep[e.g.][]{Poggianti99,Balogh02}. 
It is now clear that most such galaxies are not truly passive (in the sense of the IPS galaxies): 
instead there is evidence at 8\micron\ of some ongoing activity. 
This indicates that the morphological transformation is more intimately linked to (and concurrent with) 
the complete cessation of star formation than previously believed. 

The distribution of galaxies in the $\fratiok$-EW[OII] plane exhibits a lot of scatter
around the median relation. 
Both EW[OII] and $\fratiok$ are highly sensitive to the presence of dust 
(through extinction of the [OII] emission by dust, and abundance of PAH carriers), 
which is itself dependent on the galaxy mass and star formation rate 
\citep{Giovanelli95,Wang96,Masters03,Brinchmann04}. 
Thus whilst MIR diagnostics are better at tracing activity in high mass
galaxies, activity in lower mass galaxies may be better traced with optical diagnostics 
\citep[especially at significantly sub-solar metallicity and thus very low mass where PAH features disappear, ][]{Engelbracht05,Draine07}. 
However, within the mass range traced here, the median relation for EW[OII] goes as $\sim \Mstellar^{-0.8}$, similar to the dependence of $\fratiok$ (once corrected for the stellar atmospheric contribution). 
To compute EW[OII], the [OII] line flux is normalized by the continuum at $\lambda3727$\AA, which will lead to a shallower slope than would a normalizion by mass. 
It is likely that this effect counters any steeper relation expected due to different dust dependencies. 
Therefore, other physical parameters are necessary to drive the scatter in the $\fratiok$-EW[OII] plane, together with large measurement errors. 

There exists a strange population of galaxies extending to high EW[OII] at low $\fratiok$. 
These galaxies diverge from the usual situation in massive galaxies: Namely, that infrared data traces activity 
to much lower levels than emission lines. 
For two of the four galaxies with $\fratiok\leq0.5$ and EW[OII]$>20$\AA, the redshift depends upon a single emission line 
and might therefore be incorrect. 
Another possibility is that the [OII] emission originates in a low mass/metallicity galaxy which is falsely matched to a bright
IRAC source. Visual checks suggest this is unlikely, but cannot be certain. 
Alternatively, these galaxies might be unusually bright at 3.6\micron\ for their mass or metallicity. 
This may occur in metal-poor high mass galaxies (weird) or galaxies with a strong and very hot 
AGN component contributing most significantly to the 3.6\micron-band 
(and AGN activity can also boost the [OII] emission, \citet{Yan06}).
The only morphologically classified galaxy at high EW[OII] and low $\fratiok$ is a merger galaxy. 
This could be explained by either of the above scenarios, or a third scenario in which a merger-driven starburst produces such a hard radiation field that it is capable of destroying the PAH carriers. 
Whatever their origin, these galaxies form a puzzling, possibly interesting, but small minority. 

\section{Frequency of Infrared-Excess Massive Galaxies}\label{sec:fracIR}

Within our representative sample of $0.3\leq z\leq0.48$ and $\Mstellar \geq 2\times10^{10}\Msol$ galaxies, 
we now measure the fraction of {\bf InfraRed Excess} galaxies, $\fire$. 

Activity is traced with $\fratiok$ down to very low levels at which both star formation and nuclear activity play a role. 
We will not distinguish between these two forms of activity in this paper, instead merely computing the total fraction of IRE galaxies. 
However, any galaxy which has gas accreting onto a super-massive black hole 
is likely to be simultaneously forming stars \citep{Brinchmann04,Berta07,Shi07}. 
Hence such galaxies are correctly distinguished from the passive population. 

\begin{figure*}
  \epsscale{0.75}
  \plotone{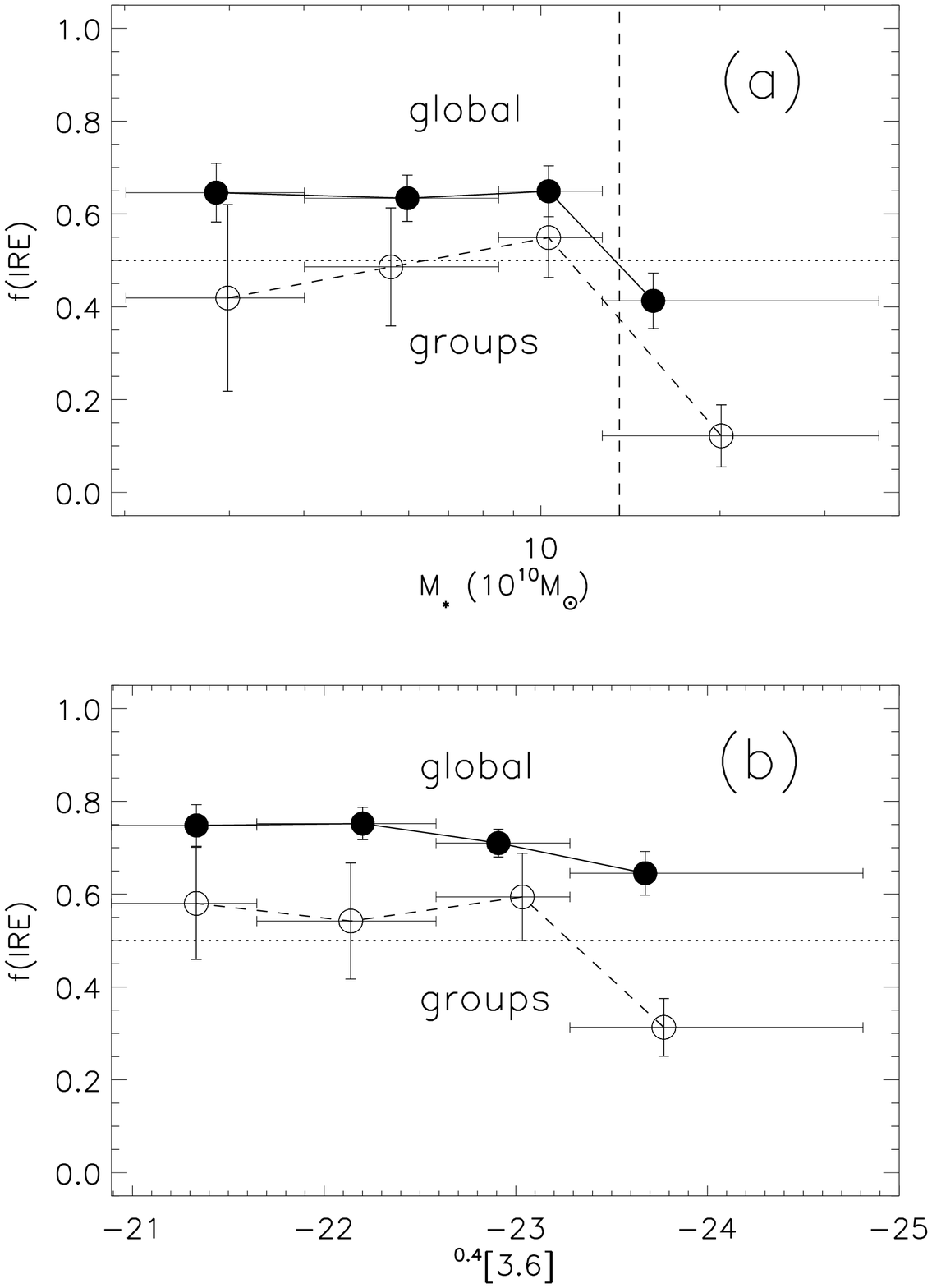}
  \caption{{\bf (a)} The dependence of $\fire$ (the fraction of InfraRed Excess
    galaxies, defined by $\fratiok>0.5$) 
    for galaxies at $0.3\leq z\leq 0.48$, 
    divided into four bins of stellar mass ($\Mstellar$). 
    This is computed for the full (global) sample (solid line, solid points) and the group galaxy sample 
    (dashed line, open points). 
    Points are plotted at the median $\Mstellar$ of galaxies in each bin, 
    and errors on $\fire$ are computed using a bootstrapping method (see text). 
    There is a significant deficit of InfraRed Excess galaxies in groups, 
    which nonetheless mirror the stellar mass dependence of the global population ($\fire$ decreases at high mass). 
    A ``\emph{crossover mass}'' ($\Mcr$) at which $\fire=0.5$ (dotted horizontal line) is defined: 
    For the global sample $\Mcr\sim1.3\times10^{11}\Msol$ (dashed vertical line).
  {\bf (b)} As (a), except $\fire$ is plotted against $^{0.4}$[3.6] luminosity for galaxies in a luminosity-limited 
  sample.}
  \label{figure:fIREMass}
\end{figure*}

To measure $\fire$, the fraction of galaxies with $\fratiok>0.5$ is computed, 
with galaxies weighted to account for the 
magnitude-dependent selection function (Section~\ref{sec:samplesel}). 
The application of this weight does not significantly change the results. 
To place errors on $\fire$, we perform bootstrap resampling of the sample. 
This is done by creating 1000 independent samples of the same size as the true sample: 
Within each sample, values of $\fratiok$ are taken randomly and often repeatedly from the true sample, 
and scattered using a gaussian kernel of width equal to the error on $\fratiok$. 
The standard deviation of $\fire$ from the bootstrap samples is taken for the error on $\fire$. 

Figure~\ref{figure:fIREMass}{\bf (a)} (solid points and line) shows $\fire$ binned by stellar mass for the global population. 
Bins are not equally populated. 
Instead they are defined such that there is an approximately equal number of group galaxies per bin (\S~\ref{sec:environment}). 
$\fratiok$ declines with $\Mstellar$, an effect seen previously in the median and percentiles. 
However this is only evident in the highest mass bin. 

To test for effects of cosmic variance, this exercise is repeated independently for each of the three CNOC2 patches with IRAC coverage
\footnote{CNOC2 consists of four patches spaced at intervals of $\sim90$\degr\ in RA, of which three are partially mapped by IRAC.}. 
The patch to patch offsets in $\fire$ are $\lesssim 3\%$,
compatible with the bootstrap errors, and indicating that cosmic
variance is not a dominant source of error. Therefore this result is
robust. 

The {\it crossover mass} ($\Mcr$, where $\fire=0.5$: dotted horizontal line), 
is $\Mcr\sim1.3\times10^{11}\Msol$ (dashed vertical line). 
\citet{Hopkins07} compiled estimates of $\Mcr$ (their ``$\rm M_{tr}$'') at different redshift. 
These are scaled to a Chabrier IMF and our cosmological parameters for comparison with our measurement. 
Defining $\Mcr$ as the stellar mass at which 50$\%$ of galaxies are optically blue or have a high specific star formation rate (calibrated using emission lines), 
they measure $\Mcr \sim 1.6-3.8\times10^{10}\Msol$ at $z\sim0.4$, 
almost an order of magnitude lower than our estimate. 
This probably reflects our sensitivity to low levels of star formation in high mass galaxies. 
However, a definition of 50$\%$ morphologically late-type galaxies results in an estimate of $\Mcr \sim 3-8\times10^{10}\Msol$ at $z\sim0.4$, 
much more compatible with our value within errors. 
This suggests that morphological selection is identifying more or less the same population as infrared selection. 
This is interesting, as it tells us that morphological transformation usually goes hand in hand with the final 
truncation of star formation in massive galaxies.

\section{The Group Environment}\label{sec:environment}

We will now examine how the fraction of InfraRed Excess galaxies depends 
upon the group environment by computing $\fire$ for group galaxies as a function of stellar mass. 
Groups were detected as overdensities in redshift space by applying a friends-of-friends algorithm 
to the CNOC2 redshift survey \citep{Carlberg01}. 
Membership has been reassigned after inclusion of additional spectroscopy and application of the membership algorithm defined in 
\citet{Wilman05}\footnote{The group aspect ratio, b, is set to 5.}. 
To ensure most group galaxies are not interlopers (i.e. nearby galaxies not bound to the group at the time of observation) 
the sample is limited to those within $500~\rm kpc$ of the luminosity-weighted group centre. 
$500~\rm kpc$ corresponds to the 
virial radius for a $\sim 360~\rm km~s^{-1}$ group at $z$=$0.4$. 
Our decision to select group galaxies within a fixed radial aperture 
instead of a scaled aperture (e.g. the virial radius) is made because 
measurements of velocity dispersions are uncertain, 
and known to be heavily biased for systems with $\lesssim 30$ members \citep[e.g.][]{Zabludoff98}. 
78/333 galaxies from the mass-selected sample match these group criteria. 
Of these $59\%$ are in groups with measured velocity 
dispersions, $\sigma_{intr}<360~\rm km~s^{-1}$; $80\%$ are 
in groups with $\sigma_{intr}<500~\rm km~s^{-1}$; and $6\%$ are in 
the cluster g226 ($\sigma_{intr}\sim 850~\rm km~s^{-1}$). 

The dashed line and open symbols in Figure~\ref{figure:fIREMass}{\bf (a)} show how 
$\fire$ depends upon stellar mass for group galaxies only. 
Each bin contains 19 or 20 group galaxies by design, 
and appears to show a deficit of IRE galaxies with respect to the global population. 
This is not highly significant in individual mass bins, but is consistent at all masses. 
Thus we estimate the combined significance for the whole sample. 
In groups, the combined value of $\fire$ is 0.447. 
As the mass distribution of group galaxies is biased to high mass relative to the global sample, 
the global sample is resampled to match the group sample on mass. 
For each group galaxy, a random galaxy is selected from the nearest 10 galaxies in stellar mass (from the global sample, other than itself, and with no repeats). 
This resampling process is repeated 10000 times, each time computing $\fire$ for the matched sample. 
On only 56 out of 10000 occasions is $\fire\leq 0.447$. 
This corresponds to a $0.56\%$ likelihood ($\sim2.5\sigma$ from a one-tailed gaussian) 
that the low value of $\fire$ in groups arises by chance. 

The lower value of $\fire$ in groups {\it independent of the stellar mass} indicates that activity 
in $\Mstellar \geq 2\times10^{10}\Msol$ galaxies is often suppressed by the group environment. 
As these groups are predominantly low velocity dispersion, often unvirialised systems, this means that
massive galaxies are already affected by the low mass halo environment. 
The mechanism of suppression seems to be particularly active in the highest stellar mass bin 
($\Mstellar\geq1.27\times10^{11}\Msol$) where in groups $\fire=0.12\pm0.07$. 
In this mass bin the group galaxies also dominate the passive global population: 
Out of a total of 43 galaxies (25 passive), 20 are in our group sample (17 passive). 
Hence the most massive galaxies seem to become passive in the group environment.
This result is even more remarkable when it is considered that the group sample is incomplete due to incompleteness effects in the CNOC2 redshift survey, and so the global sample should include more group galaxies than have been identified. 

It is prudent to repeat this exercise as a function of $^{0.4}$[3.6] luminosity 
to ensure this result does not somehow depend upon our mass calibration. 
Figure~\ref{figure:fIREMass}{\bf (b)} shows how $\fire$ depends upon $^{0.4}$[3.6] luminosity for a 
luminosity-selected sample ($^{0.4}$[3.6]$\leq-20.89$). 
Luminosity selection allows an additional 180 $\Mstellar < 2\times10^{10}\Msol$ galaxies to make it into the sample with lower than maximum mass to light ratios. 
Effectively this means that the overall value of f(IRE) is higher for the luminosity-selected sample ($0.735\pm0.018$) than for the mass-selected sample ($0.621\pm0.030$). 
The trends mirror those seen as a function of mass with the exception that in the bin of highest luminosity the global $\fire$ is still quite high (0.645$\pm$0.047). 
This is because IRE galaxies are more likely to be found in the high luminosity bin than in the high mass bin, due to their lower mass to light ratios. 
In groups the dominance of passive galaxies is so strong as to mitigate this effect. 

\section{Discussion}\label{sec:discussion}

MIR colour is a highly sensitive tracer of activity in galaxies. 
This has allowed us to separate truly passive galaxies (typically morphologically 
early-type) from galaxies with InfraRed Excess at 8\micron\ (IRE galaxies). 
In this way we avoid confusion between galaxies with low but still ongoing star formation and those 
which have shutdown their star formation altogether (probably coincident with the morphological transformation). 
Other studies examine the evolution in fractions of highly star forming galaxies \citep[LIRGs, ULIRGs etc:][]{Bell05,LeFloch05,Hammer05,Caputi06,Daddi07}. 
These fractions decline steeply with cosmic time, with contributions from generally declining star formation rates in most galaxies, and complete suppression in some. 

Our main result is that a strong and highly significant deficit in the fraction of InfraRed Excess galaxies, 
$\fire$, exists in optically selected, mostly low velocity dispersion groups at z$\sim$0.4, and is \emph{independent of stellar mass}. 
The suppression of infrared activity in low redshift Hickson Compact groups \citep{Johnson07} therefore extends 
to the ``loose group'' regime at z$\sim$0.4. 

Our group data indicate a dramatic drop in $\fire$ at stellar masses $\Mstellar\gtrsim10^{11}\Msol$. 
This coincides with the stellar mass at which a galaxy's own halo usually exceeds $\Mhalo\sim5\times10^{12}\Msol$, 
which is also the mass of the Local Group \citep{LiWhite07} and typical of the lower mass groups in our sample 
(McGee et al., submitted). 
Therefore galaxies with $\Mstellar\gtrsim10^{11}\Msol$ are typically ``central'' galaxies in groups, which is why 
many of our most massive galaxies are members of known groups. 
\emph{The low mass group environment suppresses star formation in both central massive galaxies and their satellites.  
The strongest effect is in the central galaxies.} 

In this paper, we do not fully constrain the mechanism(s) responsible for suppressing activity in groups. 
For a discussion of the menagerie of possibilities we refer to \citet[][and references therein]{Moran07}. 
The ability to detect low levels of activity has important implications for the interpretation of passive 
spirals as identified purely from their [OII] emission \citep{Poggianti99,Balogh02,Moran06}. 
Our strongest direct constraint on the mechanism(s) at work is that it (they) must be active in low mass group haloes. 
This probably excludes processes such as ram-pressure stripping which prefer much higher density environments 
\citep{GunnGott72,Quilis00}. 

Relating the suppression of star formation to a halo mass threshold is theoretically attractive, 
as the physics of gas accretion and feedback are more likely to relate to the gravitationally bound material within the halo than to the stellar mass of the galaxy itself \citep[e.g.][]{Dekel06}. 
For example, AGN feedback processes might be especially efficient in galaxies located inside group-sized haloes. 
Brightest Group Galaxies (BGGs) more frequently contain radio-loud AGN than similarly 
massive non-BGGs. 
This is likely to be related to the cooling of the hot Intra-Group Medium (IGM) onto the BGG, 
fuelling a radio-loud AGN \citep{Best07}. 

\begin{figure*}
  \epsscale{0.80}
  \plotone{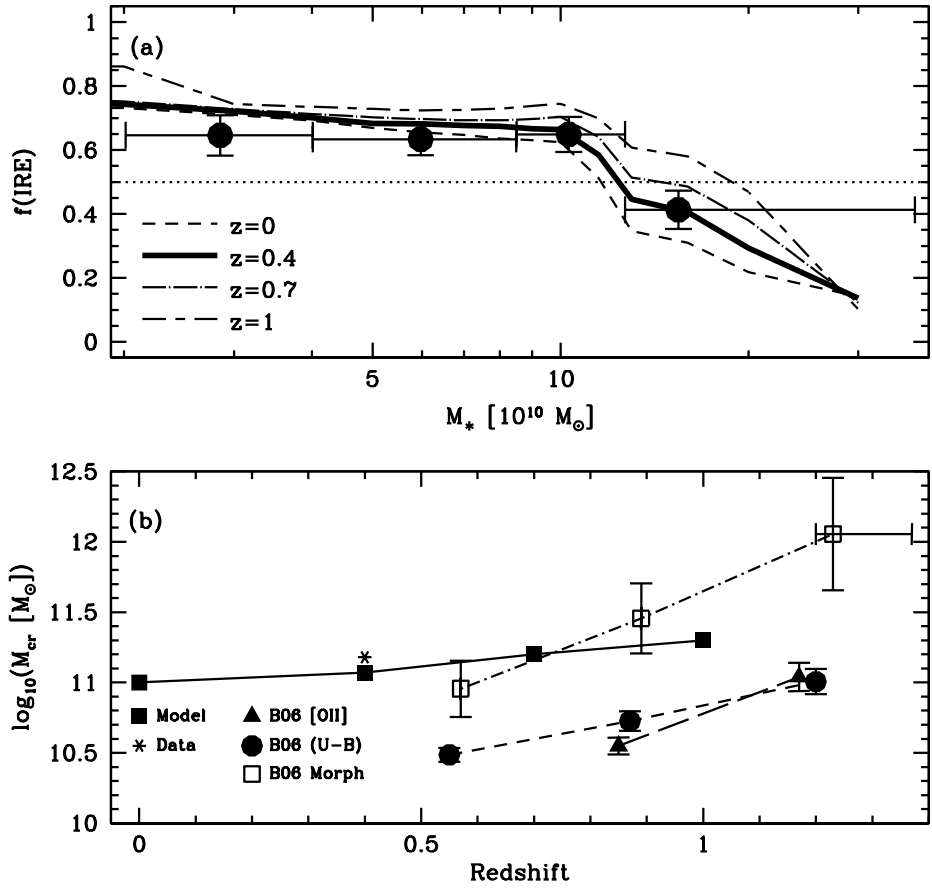}
  \caption{{\it\bf (a)} The dependence of $\fire$ on stellar mass $\Mstellar$ for the global population of $0.3\leq z\leq 0.48$ CNOC2 galaxies compared with our simple model predictions at $z=0.4$. 
    The model predictions are also computed at $z=0$, $0.7$, and $1$. 
    Galaxies in haloes with $\Mhalo\geq5\times10^{12}\Msol$ are considered to be in groups. 
    The stellar mass dependence can be reproduced by a 
    model in which suppression of galaxy activity is driven primarily by the group environment. 
    {\it\bf (b)} Evolution of the crossover mass $\Mcr$ (where $50\%$ of all galaxies are active, computed in various ways). 
    Our data point ($\Mcr$ at $z$=$0.4$ computed using $\fire=0.5$ in CNOC2) is compared with the 
    model predictions at $z=0,0.4,0.7$, and $1$, 
    and DEEP2 crossover masses computed in different redshift bins by \citet[][B06]{Bundy06} for 
    $50\%$ late-type morphologies, 
    $50\%$ galaxies blue in (U-B) colour, 
    and $50\%$ [OII]-inferred SFR above 0.2$\Msolyr$. 
    Our simple model can qualitatively reproduce a downsizing scenario, 
    although the redshift dependence is shallower than observed by B06. 
  }
  \label{figure:MillSimfIRE}
\end{figure*}

Our data alone is not sufficient to examine how the suppression of star formation within group-mass haloes contributes to the global population in a statistical manner. 
Incompleteness of the CNOC2 redshift survey and uncertainties in the construction of the group catalogue and galaxy membership inevitably leads to incompleteness of our group galaxy sample in a way which may depend upon galaxy stellar mass. 
To better estimate membership on a statistical basis, 
we use a simulated galaxy catalogue in which stellar mass and embedding halo mass can be directly evaluated for each galaxy at $z$=$0.4$. 
We choose the ``mock'' galaxy catalogue of \citet{Bower06} which is based upon the Millenium Simulation, 
a simulated $\Lambda$CDM Universe with $500h_{100}^{-1}\Mpc$ sides \citep{Springel05}. 
These dark matter haloes are populated with mock galaxies using a semi-analytic technique. 
Only the positions (environment) and stellar masses of these galaxies are used. 
The mock galaxies are assumed to be in groups where they are located within haloes $\Mhalo \geq 5\times10^{12}\Msol$. 
Mock group galaxies are then assigned a probability that they are IRE galaxies consistent with the data: 
i.e. $\fire=0.5$ for the $\Mstellar<10^{11}\Msol$ galaxies and 
$\fire=0.1$ for the $\Mstellar\geq10^{11}\Msol$ galaxies (as seen in Figure~\ref{figure:fIREMass}{\bf (a)}). 
Combined with the rest of the mock galaxy population (``field'' galaxies), 
the dependence of $\fire$ upon stellar mass is compared to the CNOC2 global trend. 
By setting the fraction of IRE field galaxies to 0.9, regardless of their stellar mass, 
the global trend is well reproduced. 
This is demonstrated in Figure~\ref{figure:MillSimfIRE} {\bf (a)} in which the observed global trend for CNOC2 galaxies 
(solid points) is extremely well matched by the model (solid line). 
That the fraction of field galaxies which need to be passive is low (0.1) {\it and independent of stellar mass} 
lends support to our hypothesis that the halo mass is more 
important than the stellar mass for the suppression of star formation in galaxies.

Figure~\ref{figure:MillSimfIRE} {\bf (a)} also shows how $\fire$ depends upon stellar mass for the mock galaxy population 
evaluated at $z=0$, $0.7$, and $1$. 
Remarkably, the model exhibits a \emph{downsizing} behaviour, such that the crossover mass $\Mcr$ decreases at lower redshifts. 
This is shown in Figure~\ref{figure:MillSimfIRE} {\bf (b)} with the CNOC2 datapoint at $z$=$0.4$. 
The only reason for this evolution is that the fraction of galaxies in groups increases, as a function of stellar mass, 
from $z=1$ to $z=0$. 
\emph{Suppression of galaxy activity in groups, plus the gravitational growth of structure in the Universe, 
can explain downsizing trends in $\Mcr$.}

Also in Figure~\ref{figure:MillSimfIRE} {\bf (b)} the crossover masses computed in different redshift bins by \citet{Bundy06} are overplotted (reproduced from their Figure~9: $\Mcr$ is called the ``\emph{transition mass}'' in that paper). 
These measurements are based on the DEEP2 redshift survey, and contribute to the compilation of \citet{Hopkins07}. 
$\Mcr$ is computed for $50\%$ late-type morphologies, 
$50\%$ galaxies blue in (U-B) colour, 
and $50\%$ galaxies with [OII]-inferred SFR above 0.2$\Msolyr$. 
That more morphologically classified late-type galaxies are detected, resulting in a higher value of $\Mcr$, 
is fully in line with our estimate at z=0.4 using a sensitive MIR colour, well correlated with morphology. 
Whilst our simple model can qualitatively reproduce a downsizing scenario, comparison with \citet{Bundy06} suggests that the 
redshift dependence is probably too shallow. 
With better constraints at different redshift, 
a model in which star formation is not immediately suppressed upon incorporation of a galaxy into a massive halo might serve to steepen this 
evolution.

\section{Conclusions}\label{sec:concl}

From the CNOC2 redshift survey, we have constructed a mass-selected sample of galaxies with known
spectroscopic redshifts $0.3\leq z \leq0.48$, and stellar masses
$\Mstellar\geq 2\times 10^{10}\Msol$. Whilst the sample is not
fully complete, the selection is well understood, 
and we correct for incompleteness using a simple weighting scheme. 
Within this sample, the fraction of galaxies with excess emission at 8\micron\ is computed, which traces star formation or nuclear activity. 
For galaxies in this sample, the Mid InfraRed (MIR) colour 
$\fratiok$ has been measured. 
This is the flux ratio of the 8\micron\ to 3.6\micron\ IRAC bands, 
k-corrected to a consistent rest-frame of a $z$=$0.4$ galaxy. 
Based upon this sample we have shown:

\begin{itemize}
\item{{\bf Infrared Passive Sequence:} Old stellar populations (such as
    found in passively evolving galaxies in which star formation has
    been suppressed) exhibit very predictable colours in the $\sim2-7$\micron\ spectral region (rest-frame), 
    tracing the stellar atmospheres of giant M-stars. 
    This produces a tight {\bf Infrared Passive
    Sequence (IPS)} in the $\fratiok$ colour. The IPS does not
    suffer from the optical red sequence degeneracy between dusty red galaxies and passive
    red galaxies. The mean IPS colour ($\fratiok=0.34$ at $\Mstellar>6\times10^{10}\Msol$) 
    is consistent with an old stellar population, and the tight scatter ($0.03-0.05$ at high mass)
    is consistent with photometric errors. This leaves little room for
    large variations in metallicity, or additional contributions at
    $\sim6$\micron\ from circumstellar or diffuse dust.} 
\item{{\bf MIR Colour Bimodality:} The $\fratiok$ distribution is bimodal, consisting of the IPS and a population 
    of galaxies with significant excess emission at 8\micron\ (InfraRed-Excess Galaxies). These two populations can be divided at 
    $\fratiok=0.5$.}
\item{{\bf Morphologies and infrared emission:} 15 out of 17
    morphologically classified early-type (elliptical or S0) 
    galaxies have $\fratiok$ colours consistent with the IPS, whilst
    no classified late-type galaxy has $\fratiok<0.5$. 
    This is consistent with known correlations at low redshift between morphology and MIR colours \citep[e.g.][]{Pahre04,Li07}.}
\item{{\bf Tracing low level activity in the MIR:}
    There is a strong correlation between $\fratiok$ and EW[OII].  However, our division
    at $\fratiok=0.5$ corresponds to a division at EW[OII]$=2.1$\AA\ for the median galaxy.
    This is below the [OII] detection limit of most intermediate to high redshift surveys. This shows that
    MIR diagnostics trace low levels of activity at $z\sim0.4$. 
    Consequently, dividing at $\fratiok=0.5$ selects a high fraction of InfraRed-Excess (IRE) galaxies, $\fire$, 
    and the crossover mass at which $\fire=0.5$, $\Mcr\sim1.3\times10^{11}\Msol$, 
    is also high with respect to optical studies.}
\item{{\bf Consequences for selection of passive spirals:} There exist
    spiral galaxies in our sample with undetected [OII] emission but
    significant MIR excess emission ($\fratiok>0.5$). 
    Thus, selection of passive spirals based upon a
    simple EW[OII] division \citep[e.g.][]{Poggianti99,Balogh02,Moran06} will
    suffer contamination by dusty, low level star formation, mostly early-type spirals.}  
\item{{\bf Strong suppression of activity for massive galaxies in
      low mass groups:} 
      In the $z\sim0.4$ optically-selected group environment there is a strong and robust (2.5$\sigma$) 
      deficit in $\fire$ compared with global values, \emph{independent of stellar mass}. 
      This is true across our range in stellar mass $\Mstellar\geq2\times10^{10}\Msol$, 
      but the suppression appears to be particularly strong at high mass 
      $\Mstellar\gtrsim10^{11}\Msol$, where the fraction of active group galaxies falls to 
      $\fire=0.12\pm0.07$.}
\item{{\bf Suppression in groups can drive the stellar mass dependence, and structure growth can drive downsizing:} 
    The global dependence of $\fire$ on stellar mass at $z$=$0.4$ can be accurately reproduced by a simple model in which suppression mainly occurs in groups, and is particularly strong for $\Mstellar\geq 10^{11}\Msol$ group galaxies. 
    Furthermore, structure growth in the Universe drives more galaxies into group-sized haloes with cosmic time. 
    This automatically leads to a \emph{downsizing} phenomenon in the global population, in which the crossover mass $\Mcr$ 
    evolves to lower masses at lower redshift.}
\end{itemize}

\acknowledgements

We thank the anonymous referee for substantial help with improving the manuscript. 
We would like to thank the CNOC2 team for allowing us access to their
unpublished data, and Dr. Eckhard Sturm for providing us with the
calibrated ISO-SWS spectrum of M82 in ASCII format. This work is based
on observations made with the Spitzer Space Telescope, which is
operated by the Jet Propulsion Laboratory, California Institute of
Technology under a contract with NASA. It is also based on
observations made with the NASA/ESA Hubble Space Telescope, at the
Space Telescope Science Institute, which is operated by the Association
of Universities for Research in Astronomy, Inc., under NASA contract
NAS 5-26555. These observations are associated with program 9895. 
DW and DP are supported by the Max Planck Society.

\end{document}